\documentclass[prb,preprint]{revtex4-1}

\usepackage{graphicx}
\usepackage{xcolor}
\usepackage{tikz}
\usepackage[RPvoltages]{circuitikz}

\usepackage{natbib}
\setcitestyle{numbers}
\setcitestyle{square}

\providecommand{\keywords}[1]
{
  \small	
  \textbf{\textit{Keywords---}} #1
}

\begin{document}

\title[Experimenting with RC and RL series circuits...]{Experimenting with RC and RL series circuits using smartphones as signal generators and oscilloscopes}

\author{Ives Torriente-García$^{1}$, Francisco M. Muñoz-Pérez$^{1}$, Arturo C. Martí$^{2*}$, Martín Monteiro$^{3}$, Juan C. Castro-Palacio$^{1}$, Juan A. Monsoriu$^{1}$}
\address{$^1$Centro de Tecnolog\'ias F\'isicas,  Universitat Polit\`ecnica de Val\`encia, 46022 Val\`encia, Spain.\\
$^2$ Instituto de F\'isica, Universidad de la Rep\'ublica, 11400 Montevideo, Uruguay. \\
$^3$ Universidad ORT Uruguay, 11100 Montevideo, Uruguay.
\\ $^*$Correspondence:  marti@fisica.edu.uy}

\begin{abstract}
Simple, portable and low-cost experiments as RC and RL series circuits are proposed  to experiment with DC circuits. Very common elements are used: a few electronics components (resistors, capacitors, coils and connecting wires) and two smartphones. We consider the charging and discharging of a capacitor in the RC  circuit and also that of coil in the RL  circuit. Using a smartphone as an oscilloscope we observe voltages variations which are  the transient response to a square signal generated in the second smartphone.
These voltage variations are directly related to the electrostatic or magnetic energy stored in the circuits.  The experimental data have  been collected with the smartphone used as an oscilloscope
and corroborated with theoretical predictions based on Kirchhoff's laws. The comparison showed  differences of the order of the 1\%  or less between the calculated capacitance or inductance compared to the manufacturer values. 
This approach which avoids the use of expensive signal generators, oscilloscopes, or any specialized hardware can be performed in less-favored contexts and even as a home assignment. 
\end{abstract}

 
\date{\today}

\maketitle 

\section{Introduction}
Although smartphones became popular only a decade ago their use in physics laboratories at all levels has become commonplace \cite{monteiro2022resource}. The experiments where they are used cover all areas of physics, but in some of them, such as mechanics or waves, it happens more often. In some areas such as electromagnetism their use has focused on experiments involving the measurement of magnetic fields
\cite{0143-0807-36-6-065002,monteiro2017magnetic,salinas2020analyzing,monteiro2020magnetic}. However, experiments involving electrical circuits have received less attention (see for example \cite{forinash2012smartphones,groff2019estimating}). In this paper we propose a set of experiments about simple circuits  using a minimal set of electronic components and two smartphones.

Direct current (DC) circuits combining resistors, capacitors and coils such as the RC series circuits or the RL series circuit are a central topic in introductory physics courses. In experimental courses these circuits are approached using signal  generators, oscilloscopes and multimeters. These instruments are not always available in all institutions, nor it is common for students to have them at home. The experiments proposed here present the advantage that they can be instrumented in less-favored contexts, but they can also be proposed with a minimum of additional materials as a home assignment for students. 

Several ingenious proposals for experiments on simple circuits can be found in the literature. Among them we can highlight \cite{pili2019using} which proposed an experiment to study the charge and discharge of a capacitor in an RC circuit using a digital voltmeter, a stopwatch and a smartphone as a high-speed camera.  Using video analysis it is possible to capture simultaneously the readings of both the voltage across the capacitor and the time elapsed. 
In another proposal \cite{groff2019estimating} a capacitor is discharged through a resistor and a buzzer. The intensity of the sound produced by the buzzer can be easily related to the voltage across the capacitor and the sound level can be used to estimate the time constant of the circuit.
Another approach to LR-circuits was recently proposed in \cite{westermann2022measuring} using the  magnetic field sensor of a smartphone to measure the magnetic field of a coil. It is also worth mentioning another experimental proposal which combines the use of video analysis an analog instruments to experiment with DC circuits \cite{aguilar2018using}.

The role of smartphones as oscilloscopes was first proposed by Forinash and Wisman \cite{forinash2012smartphones} using the headphone port designed for
connecting an external microphone and speakers to receive data from an external circuit. Another alternative to transform smartphones into oscilloscopes  takes the advantage of using Arduino boards to communicate with a smartphone using the Bluetooth interface 
\cite{ballesta2020arduino}. This approach can be used even in the case of smartphones which do not include headphone port. It is also worth mentioning the ability to transform a smartphone in an arbitrary signal generator as proposed in \cite{mathew2015mobile}.

Here, we propose an experiment which uses only a smartphone as a measuring device. Another smartphone is used as a signal generator. The outline of the article is as follows. In  Section \ref{sec:circuits} we recall the basic concepts about DC circuits. We describe the experimental setup to study  RC and RL series circuits in Section \ref{sec:setup}. In Section \ref{sec:resultsdiscussion} we present and discuss
the experimental data. Finally, we devote Section \ref{sec:conclusion} to the final remarks and perspectives.

\section{Theory on RC and RL series circuits}
\label{sec:circuits}

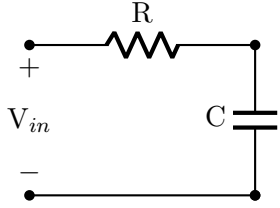
\begin{figure}[ht]
    \centering
    \begin{tikzpicture}[american,thick]
\ctikzset{
    resistors/scale=0.8,
    capacitors/scale=0.7,
}
\draw (0,0) to[short,*-*] ++ (3,0);
\draw (0,2) to[R=R,*-] ++ (3,0) coordinate(a);
\draw (a) to[C,l_=C,*-*] ++(0,-2);
\draw (0,2) to[open,v=V$_{in}$] ++(0,-2);
 \end{tikzpicture}
   \caption{Electrical diagram of a RC series circuit.}
    \label{fig:RC}
\end{figure}

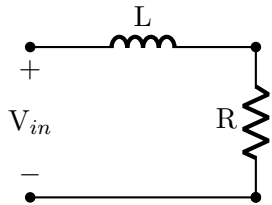
\begin{figure}[ht]
    \centering
    \begin{tikzpicture}[american,thick]
\ctikzset{
    resistors/scale=0.8,
    inductors/scale=0.7,
}
\draw (0,0) to[short,*-*] ++ (3,0);
\draw (0,2) to[L=L,*-] ++ (3,0) coordinate(a);
\draw (a) to[R,l_=R,*-*] ++(0,-2);
\draw (0,2) to[open,v=V$_{in}$] ++(0,-2);
 \end{tikzpicture}
\caption{Electrical diagram of a RL series circuit.}
    \label{fig:RL}
\end{figure}

Due to their simplicity, DC circuits are customarily introduced in introductory physics courses, in particular, they play an important role in almost all the introductory laboratories dealing with electromagnetism. The more typical examples are circuits composed by  voltage sources, resistors, capacitors, and coils disposed in a specific configuration. The first cases to be studied are usually the RC and RL series circuits, that is a resistor connected to a capacitor or coil  as shown in Figs. \ref{fig:RC} and \ref{fig:RL}.

It results convenient to study  RC or RL circuits using a
square wave as an input signal. When the input voltage is in the maximum value, the capacitor stores energy while when the input voltage is zero this energy is released.  The voltage in the capacitor is proportional to the electric field generated which in turn is readily related to electrostatic energy density. The mathematical expressions for the  voltage of  a  charging capacitor in a RC series circuit can be easily deduced using the Kirchhoff's first law
\begin{equation}
V_{\rm{C}} (t) = V_{\rm{mC}} 
(1 -e^{-t/\tau_{\rm{RC}}}),
\label{uno}
\end{equation}
where $\rm{\tau_{RC} = RC}$ is the characteristic time and $V_{\rm{mC}}$ is the maximum voltage reached at the capacitor. On the other hand, the discharging of the capacitor is expressed as,
\begin{equation}
V_{\rm{C}} (t) = V_{\rm{mC}} e^{-t/\tau_{\rm{RC}}}.
\label{dos}
\end{equation}

Analogously, using again a square wave as input signal, we will also study the charge and discharge of the RL circuit. 
While in the RC circuit we measure the voltage at the capacitor in the RL we measure the voltage at the resistor. This contrast is due to the fact that in the RC the stored energy occurs in the capacitor and the electric field is proportional to the voltage at the capacitor while in the RL circuit the energy stored in the magnetic field is proportional to the current intensity which in turn is proportional to the voltage at the resistor. The increase of the voltage at the resistor when the square wave reaches its maximum value is be expressed as
\begin{equation}
V_{\rm{R}} (t) = V_{\rm{mR}}(1-e^{-t/\tau_{\rm{RL}}}),
\label{tres}
\end{equation}
 where in this case the characteristic time is $\rm{\tau_{RL} = L/R}$ and $V_{\rm{mR}}$ is the maximum voltage reached at the resistor. Likewise, the decay of the voltage at the resistor when the square wave reaches its minimum value is expressed as
 \begin{equation}
V_{\rm{R}} (t) = V_{\rm{mR}} e^{-t/\rm{\tau_{RL}}}.
\label{cuatro}
\end{equation}
By fitting Eq.~\ref{dos} and Eq.~\ref{cuatro} to the experimental data of the voltage at the capacitor (RC circuit) during its discharging and to the experimental data of the voltage at the resistor (RL circuit), the time constants ($\tau_{RC}$ and $\tau_{RL}$) can be determined and used to calculate the capacitance and inductance, respectively.

\section{Experiments}
\label{sec:setup}

The experimental setup is composed by a simple circuit, as mentioned in the previous section a RC or a RL and two smartphones Xiaomi Redmi Note 7 and 11. Concerning the smartphone used as an oscilloscope, in previous attempts several  models were tried but not all performed  well probably due to the existence of high-pass audio filters.  One acts as a signal generator and the other as an oscilloscope. The signal generator is achieved using the tone generator provided by the free Android app Physics Toolbox Suite \cite{vieyra} . The other smartphone was used as an oscilloscope by means of free Android app phyphox \cite{staacks_2018}.
Smartphone data that can be analyzed directly on the smartphone or downloaded to a computer using the aforementioned apps.

\begin{figure}[ht]
    \centering
    \includegraphics[width=.8\columnwidth]{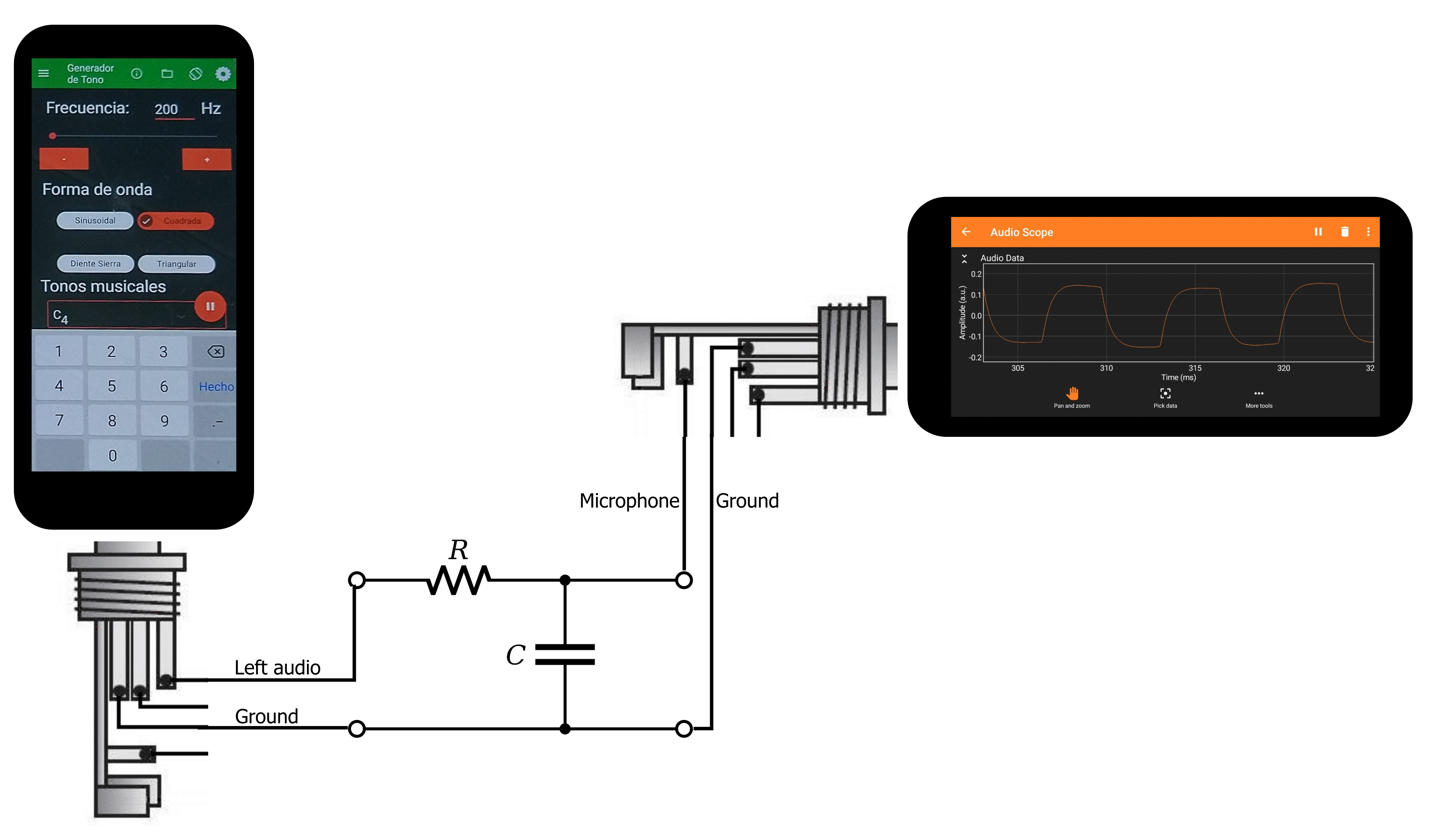}
    \caption{Schematic representation of the experimental setup to study RC circuits used in this work. In the center the RC circuit connected to the smartphones using  two dismantled TRRS jack connectors as indicated.
    The smartphone of the left is used as a signal generator using the Tone generator function of the Physics Toolbox app (in this case a
    square signal of frequency 200 Hz). In the right we observe the smartphone used as an oscilloscope by means of the   Audio amplitude option of the phyphox app.}
    \label{fig:montajeRC}
\end{figure}

\begin{figure}[ht]
    \centering
    \includegraphics[width=.8\columnwidth]{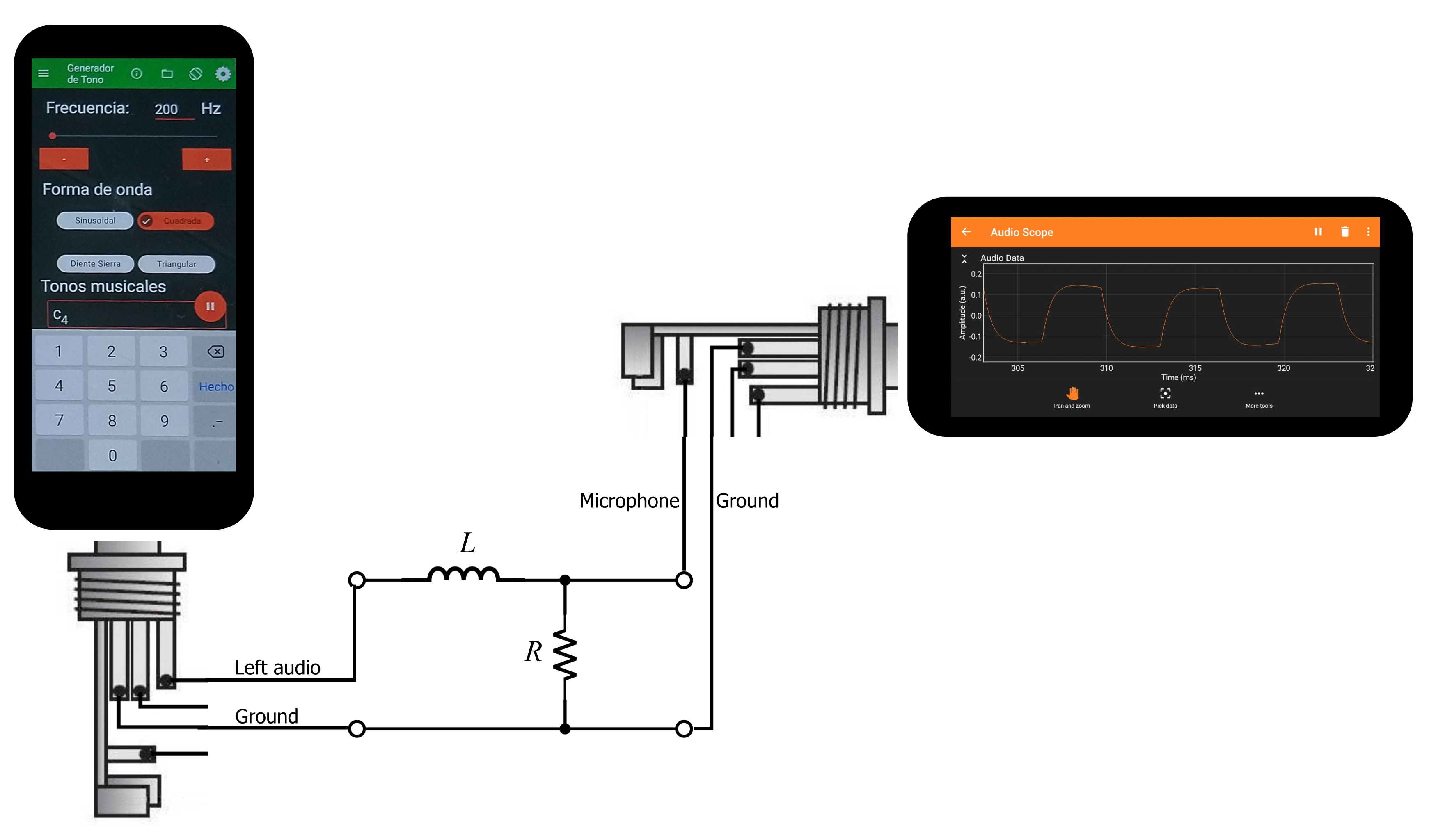}
    \caption{Experimental setup corresponding to the RL circuit studied, for details see the previous figure. }
    \label{fig:montajeRL}
\end{figure}

The charging and discharging of the capacitor (Figs. \ref{fig:RC} and \ref{fig:montajeRC}) is performed by using a square signal whose period contains several time constants. The same idea is used for the RL series circuit (Figs. \ref{fig:RL} and \ref{fig:montajeRL}).
Figures \ref{fig:montajeRC} and \ref{fig:montajeRL} (smartphones on the left-hand side of the figures) show screenshots of the tone generator option of Physics Sensors Toolbox Suite app which enables the smartphone as a signal generator. This option can be found on the left panel of the main screen of the app. Several parameters can be changed, such as the frequency, the type of signal (sinusoidal, square, saw-tooth, triangular), and the amplitude (controlled by just changing the volume level of the phone). In this experiment
we used a square signal and measured the maximum value of voltage at the audio output of 0.77 V (for the volume level set at the maximum). In the right hand-side 
of the same figures we show screenshots of the smartphones displaying the Audio Amplitude option of phyphox app which converts the smartphone to an oscilloscope.  This option is useful to visualize any audio signal sent in using the jack connector or specifically,  in this work, the voltage signals at the capacitor (RC circuit) or the resistor (RL circuit).

\begin{figure}[ht]
    \centering
    \includegraphics[width=.385\columnwidth]{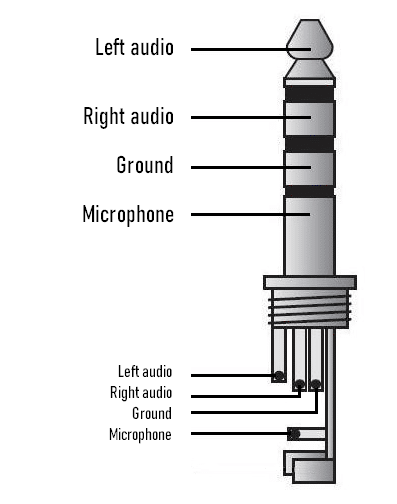}
    \caption{TRRS four-conductor jack connector and wiring diagram.}
    \label{fig:connector}
\end{figure}

An important aspect is how to connect the different elements securely. It is possible to make the connections using a  TRRS  four-conductor  jack connector readily available in electronics stores or even by disassembling an old audio cable. Figure \ref{fig:connector} shows a TRRS jack connector and the connections employed as used in this work. The connector plugged into the smartphone used as a signal generator allows powering the RC circuit while the other connector, plugged into the  smartphone used as an oscilloscope, allows acquiring the signal. For the connection of the jacks to the RC and RC series circuits, standard cables ending in crocodile (or alligator) clips have been used.
Observe that the left speaker contact of the first smartphone is  connected to the resistor, the ground cables of each smartphone are connected to the capacitor while the microphone terminal connects the second smartphone to the junction of the capacitor and the resistor.

\section{Results and discussion}
\label{sec:resultsdiscussion}
In this section we show and discuss the results obtained thanks to the smartphone acting as an oscilloscope using the phyphox app as described in the previous section. In general, voltages registered with this app are expressed in arbitrary units while time is registered in milliseconds. Voltage units are arbitrary because  the amplification factor due to the electronic circuits in the smartphone is not known a priori. Nevertheless,  as only the relative variations in the voltages are relevant,   it is still  possible to obtain  circuit time constants ($\tau_{RC}$ and $\tau_{RL}$) by means of a fitting of Eq.~\ref{dos} and Eq.~\ref{cuatro} to the collected data. 
 \begin{figure}[ht]
    \centering
    \includegraphics[width=.75\columnwidth]{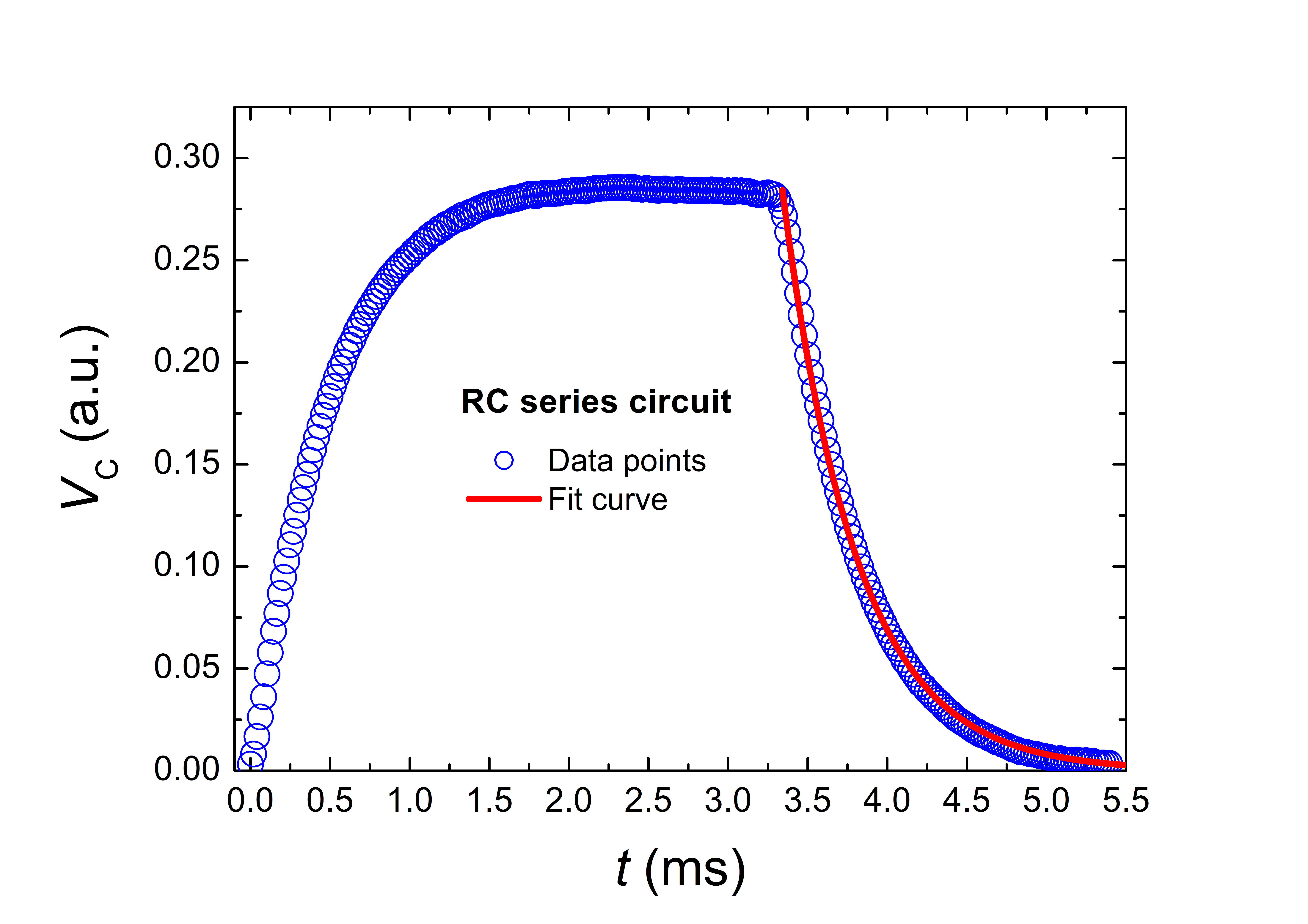}
    \caption{Voltage at the capacitor in the RC series circuit as a function of time.}
    \label{fig:plotRC}
\end{figure}

Figure \ref{fig:plotRC} shows the voltage as a function of time at the capacitor in the RC series circuit (Fig.~\ref{fig:RC}) for charging and discharging of the capacitor. The red line indicates the fit curve of the data using Eq.~\ref{dos}. The resulting characteristic time from the fitting is $\tau_{\rm{RC}}=0.465(2)$ ms. Using this value and the manufacturer value for the resistance  $R=470(24)$ $\Omega$ (manufacturing tolerance 5\%) the capacitance can be obtained as $C=\tau_{\rm{RC}}/R=0.99(5)$ $\mu$F. The manufacturer value for the capacitance is 1.0(1) $\mu$F. Therefore, the agreement between the value calculated from the experimental measurements and that provided by the manufacturer is remarkable.

\begin{figure}[ht]
    \centering
    \includegraphics[width=.75\columnwidth]{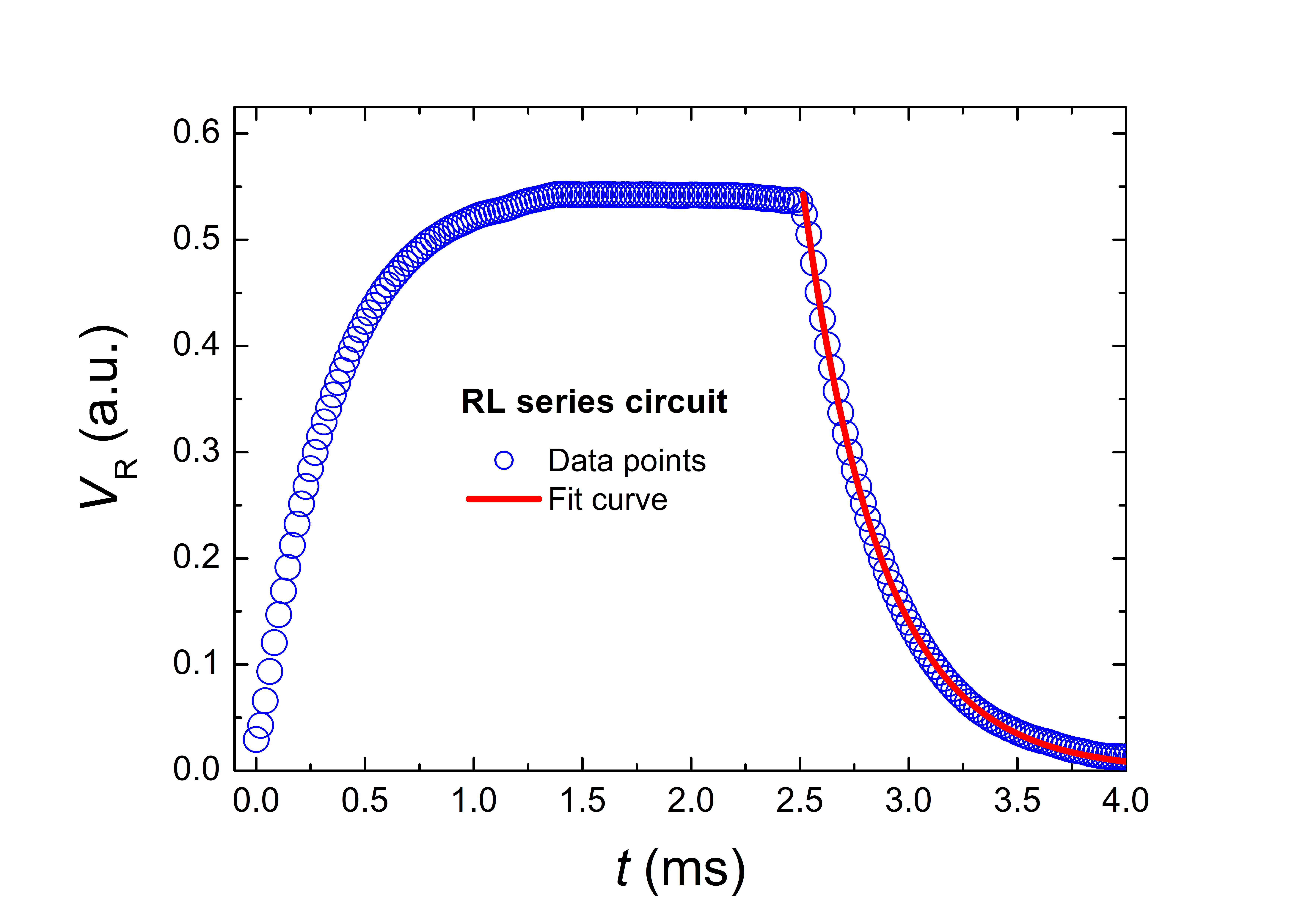}
    \caption{Voltage at the resistor in the RL series circuit as a function of time.}
    \label{fig:plotRL}
\end{figure}

Similarly, Fig. \ref{fig:plotRL} shows the voltage \textit{versus} time at the resistor in the RL series circuit (Fig.~\ref{fig:RL}) for the cases when the circuit is switched on and off, respectively. The behavior of the voltage at the resistor is linear with the behavior of the current in the circuit as $V_{\rm{R}} (t)=I(t)R$. The red line indicates the fit curve of the data using Eq.~\ref{cuatro}. 
The resulting characteristic time from the fitting is $\tau_{\rm{RL}}=0.359 (2)$ ms. Combining this value and the manufacturer value for the resistance  $R=275 (14)$ $\Omega$ the inductance can be obtained as $L=\tau_{\rm{RL}}R=0.099 (9)$ H. The manufacturer value for the inductance is 0.10(1) H. Therefore, a percentage deviation of 1 $\%$ between the calculated and manufacturer values is obtained.

 \section{Final remarks and perspectives}
\label{sec:conclusion}

We have proposed an experiment to investigate simple DC circuits and in particular RC and RL series circuits. The only elements needed for our experiments were common electronic components (resistors, coils and capacitors), connecting wires, a coupled of dismantled jack connectors and two smartphones, one used as a signal generator and the other as an oscilloscope. It is also worth highlighting that, in contrast to other approaches like \cite{Ramos_2020}, our experimental setup does not require any additional hardware or specialized electronic device.

Results for the charging and discharging of a capacitor in an RC series circuit yield a 1$\%$ of percentage difference when the calculated value of capacitance (obtained from the fitted time constant) is compared with the manufacturer value. As for the RL series constant, the inductance of the coil was calculated from the corresponding time constant, yielding a 1$\%$ of percentage deviation with respect to the manufacturer value. 

This proposal has the advantage of being portable and low-cost as it requires a minimum set of materials. It can be proposed in a wide range of socio-cultural contexts and can even be proposed to students as a challenge to be implemented at home if they have only a minimal set of electronic components and their own smartphones.

The use of smartphones as a signal generator with possibility of choosing among different types of signals (i.e. square, triangular, sinusoidal, saw-tooth) and also as an oscilloscope to visualize variable voltages, opens up the scenario for a variety of teaching innovations on this topic. For instance, RLC circuits operating with sinusoidal signals as input can be studied. 
Another possible extension could be to calibrate the signal measured by the smartphone so as to obtain a voltage expressed in S.I. units instead of arbitrary units.  
Similarly, other components such as diodes when incorporated to simple circuits (e.g. low- and high-pass filters) can be addressed as well with this approach.
  
\section{Acknowledgments}
\label{sec:agrad}
J.A.M. and J.C.C.P. would like to thank the Instituto de Ciencias de la Educación (Institute of Education Sciences) at the Universitat Politècnica de València (UPV) for its support to the teaching innovation group MSEL.
A.C.M. and M.M.  would like to thank PEDECIBA (MEC, UdelaR, Uruguay).


\bibliography{bibliography.bib}

\end{document}